# A novel mechanism for outbursts of Comet 17P/Holmes and other short-period comets

Richard Miles*

*Golden Hill Observatory, Stourton Caundle, Dorset DT10 2JP, UK*

*E-mail: rmiles@baa.u-net.com

**ABSTRACT**

A mechanism is proposed to explain the outburst of comet 17P/Holmes based on; (a) oxidation of water within the porous surface of the comet nucleus to form hydrogen peroxide ($H_2O_2$) through exposure to UV radiation, to energetic solar-wind particles and to cosmic radiation, (b) concentration of the $H_2O_2$ component through solid-, liquid- and gas-phase processes involving sublimation, evaporation, fractional crystallization, diffusion, supercooling, capillary wetting and migration in voids within the nucleus, and (c) rapid exothermic decomposition of aqueous $H_2O_2$ liberating oxygen gas via a surface catalytic reaction through interaction with finely-dispersed transition metals, metal compounds and minerals, in particular those containing Fe, localised within a differentiated multi-component comet nucleus. An accelerated release of gaseous oxygen, concomitant self-heating and volatilisation of hydrocarbons within the nucleus results in its explosive disruption. This mechanism may also explain the observation of a repeat outburst of this comet in 1893. Laboratory studies to investigate $H_2O_2$ formation in simulated cometary environments and to evaluate $H_2O_2$ decomposition on meteoritic samples are recommended.

## 1 INTRODUCTION

On 2007 October 23.7, comet 17P/Holmes began an outburst of unprecedented intensity, brightening by some 13-14 magnitudes within the space of about 24 hours thereby attaining $2^{nd}$ magnitude (CBET 1111, 1118). Amateur astronomer, J.A. Henríquez discovered the outburst on October 24.0, by which time the comet had brightened by some 6 magnitudes. It was reported that during the brightening phase, parent fragments appeared to separate from the nucleus releasing large quantities of dusty material, with a dominant event occurring around October 24.40, which was possibly responsible for the final brightness rise of the comet and the emission of a dust cloud, the outer region of which was observed to expand at about 0.5 km s$^{-1}$ (CBET 1123).

Comet 17P/Holmes is an example of a small number of Jupiter-family comets with periods of about 10 yr or less which have exhibited outbursts in which the brightness of the coma rises by 9 magnitudes or more. Others include 41P/Tuttle-Giacobini-Kresák in 1973 (Kresák 1974), 98P/Takamizawa in 1984 (http://cometography.com/pcomets/098p.html), and 97P/Metcalf-Brewington in 1991 (Kidger 1993). However, by comparison, the recent rise in brightness of 17P is some 100 times greater than has been observed in these other cases. Furthermore, the comet is known to have undergone a large outburst in 1892 November when it was first discovered by Edwin Holmes as a $4^{th}$ magnitude object having a coma about 5 arcmin across and easily visible to the naked eye. It subsequently faded but then in 1893 January it underwent a second outburst just reaching naked-eye visibility. The comet was observed at its return to perihelion in 1899 as a $16^{th}$ magnitude object but after 1906 it was lost until recovered in 1964 as an 18th magnitude object by E. Roemer based on a prediction by B.G. Marsden. Since that time, the comet has been no brighter than about 14th magnitude.

Striking similarities exist between the 1892 event and the recent one in that in both cases the comet underwent the initial outburst 5-6 months after reaching perihelion, and both events resulted in a similar evolution in appearance of the coma and tail during the following few weeks. It should also be noted that having a 7-yr orbital period, the comet has made many such perihelion passages since 1892 with no repetition of a similar outburst having been observed.

This paper describes the basis for a tentative hypothesis proposed by the author (Miles 2007) to explain these comet outbursts. It is suggested that the body of the comet is heterogeneous on a large scale being composed of a water-rich mantle and loosely-packed agglomerates rich in metal (especially Fe) compounds buried within the nucleus. The slow formation and subsequent rapid catalytic decomposition of hydrogen peroxide ($H_2O_2$) is postulated to account for the very energetic outbursts of comet 17P/Holmes and some other short-period comets.

## 1.1 Outburst models

Given the current models for comet outbursts, it is difficult if not impossible to explain the magnitude of the observed phenomena involving 17P. One early proposed mechanism (Whitney 1955) involved the vaporisation of methane or methane and carbon dioxide within contained volumes within the nucleus leading to a build up of internal pressure and finally to physical disruption. Donn & Urey (1956) proposed that the more prominent activity arises from explosive chemical reactions involving free radicals or unstable molecules, and that comet nuclei are a mixture of icy conglomerates and reactive species. It was suggested that radical species such as CH, NH and OH might form reactive compounds such as azides or $H_2O_2$, the latter being considered as an oxidant for a reaction with $CH_3OH$ and triggered by solar particle radiation. Rettig et al. (1992) investigated the possibility that HCN polymerization can provide the energy needed for outbursts and concluded that in principle it could, given an inhomogeneous nucleus and the possibility that polymers so formed would react with water to produce $CO_2$ and CO with a resultant pressure build-up in the interior. Prialnik & Podolak (1999) modelled radioactive heating and the build-up of gas pressure within the icy interior as a process leading to the break up of fragile nuclei. Totha & Lisseb (2006) considered rotational break-up of comet nuclei and of centaurs through centrifugal forces, concluding however that such events would be very rare. A speculative suggestion by Seitzl et al. (2006) involves the slow build-up of lattice defects through radiation damage of crystalline materials at very low temperatures resulting in energy densities of up to several MJ $kg^{-1}$ in some materials: the energy is released on warming when solid-state defects recombine creating the outburst.

Gronkowski (2002; 2007a) has reviewed the potential sources of energy of comet outbursts. He examines various hypotheses include the more recent ones of; (a) destruction of cometary grains in the field of strong solar wind, and (b) the transformation of amorphous water ice into the crystalline allotrope. One main conclusion of Gronkowski's work is that comet outbursts can have different causes although he suggests (2007a) that the hypothesis concerning the amorphous water ice transformation appears to be the most probable one. The concept of destruction of cometary grains through inclusion of volatile species within the grains is further developed by Gronkowski (2007b).

Whipple (1984) interpreted the two outbursts of 17P in 1892-3 as being consistent with the grazing encounter of a small natural satellite with the nucleus on November 4 followed by its final encounter in 1893 January. Given the unprecedented magnitude of 17P's recent outburst, it seems highly improbable that a further satellite has by chance collided with the nucleus. Indeed, the concept of impact by an orbiting fragment or by a passing meteoroid has been examined by Gronkowski (2004) for the case of the multiple outbursts exhibited by comet 29P/Schwassmann-Wachmann. He has shown that the probability of such events is also extremely low.

Belton et al. (2007) propose a new model for the interior of Jupiter-family comet nuclei, called the talps or "layered pile" model, in which the interior consists of a core overlain by a pile of randomly stacked layers. These layers may have accumulated over very long times whilst the precursor comet was formed in the outer solar nebula. They speculate that outbursts could have multiple physical causes in the talps model all related to a compositional inhomogeneity such as the local enhancement of a super volatile (e.g. $CO_2$, CO).

From the foregoing we may assume that comet outbursts arise via more than a single mechanism. For instance, a different mechanism may operate if the comet is in a hyperbolic orbit and is making its first close approach to the Sun, compared to the case of Centaurs such as 29P/Schwassmann-Wachmann or 2060 Chiron, which inhabit a much colder environment. By comparison, 17P is periodic and intermediate in position relative to the Sun having a perihelion distance, $q = 2.05$ AU, and eccentricity, $e = 0.432$. Although many comets share similar orbital characteristics to 17P, of the 23 periodic comets for which $1.75<q<2.35$ AU and $0.35<e<0.55$ for example, there are none which exhibit significant outbursts, indicating that the nature of 17P must be special.

## 2 ROLE OF HYDROGEN PEROXIDE IN COMET OUTBURSTS

### 2.1 Formation of hydrogen peroxide

Water is a major constituent of many comets as evidenced for example by observations of the neutral hydrogen coma in comet C/1999 S4 (LINEAR) made with the Lyman-alpha imager of the SWAN instrument on the SOHO spacecraft (Mäkinen et al. 2001). When comets are more than about 3 AU from the Sun, water exists largely in the form of ice. The immediate surface is subject to bombardment by high energy nuclei and electrons from the Sun, which affect the upper few meters of the mantle as well as being subject to cosmic radiation, which can penetrate to greater depth. We know from Galileo spacecraft observations of Europa, that $H_2O_2$ is formed at concentrations up to 0.3 mol% on its ice-rich surface through irradiation (Carlson et al. 1999, Hendrix et al. 1999). Hendrix et al. 1999 also found $H_2O_2$ on Ganymede and Callisto. It has also been detected on Mars and its distribution and seasonal variations mapped (Encrenaz et al. 2004). Most recently, Cassini spacecraft observations have revealed the existence of $H_2O_2$ on Saturn's moon, Enceladus (Newman et al. 2007).

Laboratory studies (Loeffler & Baragiola 2005) have shown that under the conditions on Europa at a temperature of 80 °K, hydrogen peroxide molecules exist in isolated trimers $H_2O_2 \cdot 2H_2O$, and if the temperature is raised to 150 °K the trimers become mobile and aggregate to form precipitates of crystalline dihydrate $H_2O_2 \cdot 2H_2O$. Laboratory experiments by various workers have shown that $H_2O_2$ is the major product (along with hydrogen, which readily escapes) when water ice is irradiated with energetic electrons (Zheng, Jewitt & Kaiser 2006) or atomic ions $H^+$, $He^{++}$ and $Ar^{++}$ (Gomis, Leto & Strazzulla 2004). Zheng et al. (2006) were able to demonstrate the formation of up to 3.3 mol% $H_2O_2$ in their experiments. These studies simulated conditions in the outer solar system or in cold molecular clouds in interstellar space. They did not however come close to reproducing conditions near the surface of a comet nucleus at heliocentric radii of 2.0-2.5 AU for example. On icy surfaces at low temperature, the concentration of $H_2O_2$ will reach a limiting value when the rate of formation equals the rate of destruction by the incident radiation. In contrast, where comets approach closer to the Sun, melt water can flow in the immediate subsurface such that UV radiation and energetic particles can react to form $H_2O_2$, which migrates or diffuses away into the bulk liquid where it remains. Formation of $H_2O_2$ by photolysis of water by UV radiation is well known. Indeed, given the O-H bond energy of the $H_2O$ molecule is close to 459 kJ mol$^{-1}$ (http://www.lsbu.ac.uk/water/data.html), UV radiation short of about 240 nm is expected to cleave the molecule leading to the production of $H_2O_2$. A striking example of this is the aqueous phase photochemical formation of $H_2O_2$ in clouds in the Earth's atmosphere. For instance, Anastasio, Faust and Allen (1994) observed formation rates in clouds of up to 3.0 $\mu M$ h$^{-1}$. From the foregoing, it is to be expected that some short-period comets (P<10 yr) with water-rich mantles will tend to accumulate $H_2O_2$ within their water/ice matrix.

**2.2  Physical processes concentrating hydrogen peroxide within the nucleus**

The subsurface of a comet nucleus having a low albedo and rotating slowly (<1 rev/d) will generally undergo repeated heating of the sunlit face followed by subsequent cooling when the same surface region is turned away from the Sun. The lower the albedo and the slower rotation, the more the subsurface will be heated prior to the temporary cooling phase. In extreme cases, comets may heat up sufficiently that liquid water may coexist with ice in the subsurface layer at distances well beyond 2.5 AU. The effect of this diurnal heating and cooling will be to transport $H_2O_2$ formed near the surface to deeper within the nucleus. If a comet is rich in water, several physical processes including diffusion and physical mixing of the water can come into play, the result of which is to gradually increase the $H_2O_2$ concentration within localised regions of the nucleus.

Let us examine the physical properties[1] of $H_2O_2$ and $H_2O$ which might play an important role in the cometary environment:

First consider the loss of water molecules by evaporation to space, which takes place at the immediate surface and within fissures across the surface. Here, direct exposure to the vacuum of space can lead to the formation of superficial ice. However, much of the surface will comprise a porous dusty layer, the pores within which can be wetted by liquid water. Some regions of the surface are more active than others for example at fractures releasing water vapour, other volatiles and entrained dust into space by way of jets.

Importantly, the vapour pressure of $H_2O_2$ is significantly lower than that of $H_2O$ both in its pure form and in aqueous solution. For example at 273 ºK, 50-%wt aqueous $H_2O_2$ exhibits a total vapour pressure of about 250 N m$^{-2}$ whereas the partial pressure of $H_2O_2$ in the vapour phase is less than 10 N m$^{-2}$, the remainder comprising the $H_2O$ fraction. Boiling at reduced pressure is therefore a very efficient process for concentrating aqueous $H_2O_2$ : indeed vacuum distillation is used industrially to do just this (www.nzic.org.nz/ChemProcesses/production/1E.pdf). Hence $H_2O_2$ formed near the surface will tend to be concentrated by loss of the water component to space, and the peroxide so formed will diffuse away from the surface layer to be replaced by more water from the bulk. This process will lead to the accumulation of $H_2O_2$ especially if small reservoirs are present below the surface of the nucleus.

A further physical process involving melting and freezing may also concentrate the $H_2O_2$ present. Above 273 ºK, $H_2O_2$ is miscible with water in all proportions. Below this temperature, the composition and melting characteristics of bulk aqueous mixtures vary according to the solid-liquid phase diagram shown in Fig.1.

---

[1] The website at http://www.h2o2.com/intro/properties.html comprises an extensive collection of physical, thermodynamic and optical properties of $H_2O_2$ and its mixtures with $H_2O$

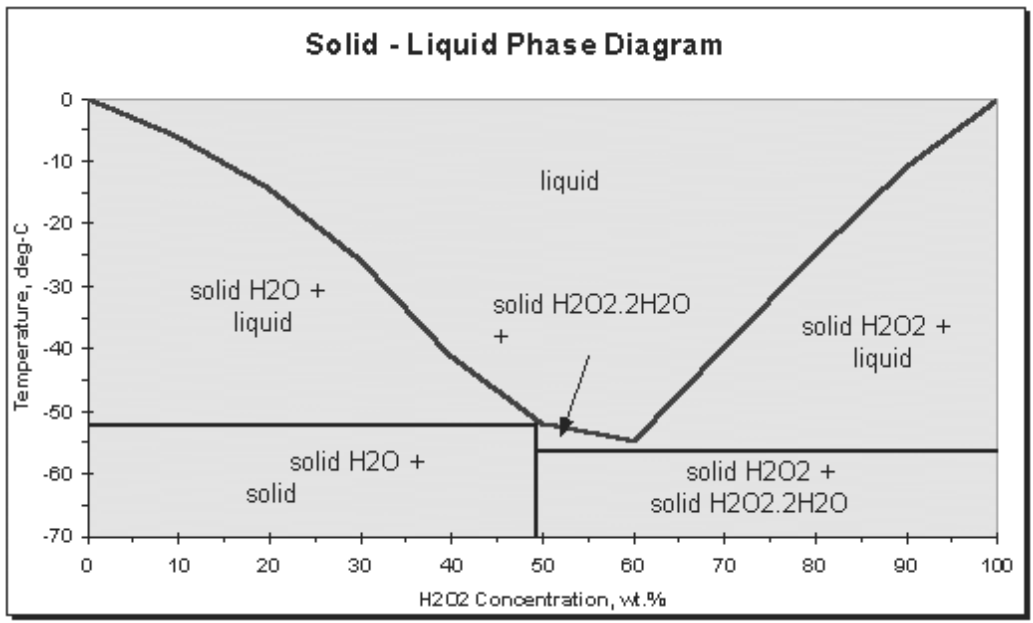

Figure 1 - Solid-liquid phase diagram for hydrogen peroxide and water.
(*from Giguère P.A., "Complements au Nouveau Traite de Chimie Minérale, No.4",
"Peroxyde d'Hydrogène et Polyoxydes d'Hydrogène", Paris, Masson 1975*)

If for instance a 20-%wt aqueous $H_2O_2$ solution is cooled, pure ice will crystallize out thereby concentrating the remaining liquor in $H_2O_2$. This process leads to the fractional crystallisation of water ice resulting in a local enhancement in the peroxide concentration - importantly this may take place once per rotation (heating/cooling cycle) of the nucleus and in a more progressive fashion on each occasion the comet recedes towards aphelion. Also as the aqueous peroxide concentration increases, so does its density. Pure $H_2O_2$ has a density of about 1.49 g cm$^{-3}$. A 20-%wt aqueous solution at 273 °K has a density of 1.08 g cm$^{-3}$. Water ice on the other hand has a density of 0.915 g cm$^{-3}$. So if the comet nucleus rotates very slowly and spends many days or weeks with one face sunlit and the opposite face hidden from sunlight, the dark side will undergo a very significant cooling such that water ice will tend to crystallize out from the liquid phase near the outer surface of the nucleus leaving the deeper-lying liquid phase enriched in $H_2O_2$. Some of the water ice exposed to the vacuum of space will also tend to be lost through sublimation.

Aqueous peroxide solution, if sufficiently concentrated, can be cooled to as low as -56ºC (217 ºK) before it freezes. Thus, as the peroxide concentration increases, the temperature range over which liquid processes can operate is also extended. Conditions within voids within the nucleus may favour the build-up of relatively high concentrations of $H_2O_2$. Such solutions are known to exhibit supercooling in the bulk phase (Cooper & Watkinson 1957) which would further assist transport at low temperature helping to fill voids within the nucleus. As the nucleus rotates, reservoirs near the surface rich in $H_2O_2$ would undergo continuous freezing and melting accompanied by volume changes which may also lead to localised fracturing of the matrix.

One other physical characteristic of comet nuclei which is important involves microporosity. Carbonaceous chondrites, possibly the nearest analogues to comet nuclei that we have, are known to be very porous. For instance, one of the most porous examples is the Orgueil CI1 meteorite for which Consolmagno & Britt (1998) measured a porosity of 35±3 %. A consequence of microporosity is capillary wetting. As the nucleus rotates, part of the surface will not be sunlit and the liquid in the immediate subsurface will freeze. Water vapour will gradually sublime into space so that the pores near the surface are emptied. On returning to the sunlit side, ice deeper within the surface will melt and capillary attraction will rapidly fill the empty pores once again bringing liquid close to the surface where $H_2O_2$ production can resume. One other characteristic of micropores is that the vapour pressure of liquids within the pores is depressed according to the Gibbs-Thomson equation, one form of which is:

$$p/p_0 = \exp(2\gamma \cdot v_{atom}) / (kT \cdot r)$$

where p is the effective vapour pressure, $p_0$ the vapour pressure in the bulk, $\gamma$ the surface tension, $v_{atom}$ the atomic volume, k Boltzmann's constant, T the absolute temperature and r the radius of the pore (http://en.wikipedia.org/wiki/Gibbs-Thomson_effect). As a result, the vapour loss from liquids within highly-porous structures exposed to the vacuum of space will be reduced. Mautner (1999) measured a value of 37 $m^2$ $g^{-1}$ for the specific surface area of the Murchison CM2 meteorite (porosity = 16±2 %, Moore, Flynn & Klock (1999)) indicative of an average pore size of about 0.01 μm. Water loss into space by evaporation from pores having dimensions of this order and smaller will therefore be much reduced. A further effect associated with liquid in micropores, also predicted by the Gibbs-Thomson equation, is melting point depression. If pore size is small, the liquid within a pore will remain liquid as it is cooled below the normal freezing point. This effect has been observed in the case of aqueous $H_2O_2$ (e.g. Li, Ross & Benham 1991, Webber & Dore 2004).

Note that the effects of supercooling, melting point depression, capillary wetting and vapour pressure lowering all conspire to maintain water in the liquid state in the immediate subsurface of comets such that it is then more easily exposed in its liquid state to UV radiation and high-energy particles from the solar wind.

The overall effect of the physical processes described above will be to enhance the aqueous concentration of $H_2O_2$ as it is formed, and to transport $H_2O_2$ from near the surface to deeper within the nucleus thus tending to form localised reservoirs within voids. How far such processes progress for any particular comet nucleus will depend on many factors. However, the rate-limiting step may not be governed by physical factors but instead by the chemical composition of the nucleus since $H_2O_2$ is potentially very reactive. For example, it decomposes thermally if heated or if aqueous solutions are made strongly alkaline. In relation to comets, its most important chemical property is its tendency to decompose through contact with solid catalysts or dissolved metal ions.

**2.3 Catalytic decomposition and the role of hydrogen peroxide**

Aqueous solutions of $H_2O_2$ readily decompose through contact with a variety of catalysts including dispersed metals, alloys and transition metal compounds (Abbot & Brown 2004, Teel et al. 2007) liberating oxygen gas according to the equation:

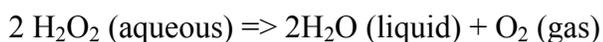
2 $H_2O_2$ (aqueous) => 2$H_2O$ (liquid) + $O_2$ (gas)

Complete decomposition of a 20-%wt aqueous $H_2O_2$ solution liberates 60 times its own volume in the form of oxygen gas at 1 atm and 293 ºK (STP). The reaction is strongly exothermic liberating 98 kJ $molH_2O_2^{-1}$ (Schumb, Satterfield & Wentworth 1955). Even a 10-%wt solution can boil at STP through self-heating if it becomes grossly contaminated, giving off large volumes of gaseous oxygen in the process (http://www.h2o2.com/intro/properties/thermodynamic.html). In contrast to its thermal decomposition, which is governed by first-order kinetics and slows in rate by a factor of 2.3 for each 10 ºK the temperature is lowered, the catalysed reaction exhibits quite different characteristics. For example, exposed to goethite (FeOOH) the rate is governed by the available surface area of FeOOH and exhibits only about one third the normal temperature dependency (Lin & Gurol 1998) with an activation energy for the reaction at the iron oxide surface of 32.8 kJ $molH_2O_2^{-1}$. Thus provided materials are catalytically active and finely-dispersed, the reaction can proceed at a fast rate even at low temperature.

Not only are iron oxides and hydroxides (hematite, goethite, etc.) surface catalysts for the decomposition of $H_2O_2$, such compounds are common in meteorites and comets as evidenced by results from the Stardust mission to comet 81P/Wild (Grossemy et al. 2007) and often exist in very finely-divided form having been assembled from interstellar grains. For most comets, the iron and other transition-metal compounds will most probably be distributed throughout the body of the comet nucleus and so will tend to decompose $H_2O_2$ as and when it is formed. However, if such compounds exist in the form of localised agglomerates buried well within the nucleus, peroxide concentrations may gradually build up to significant levels (say >5 %wt) in the subsurface over several perihelion passages without any significant tendency for the $H_2O_2$ to decompose.

One other relevant chemical characteristic of $H_2O_2$ is its ability to oxidise hydrocarbons and other organic compounds. For example, Mautner (1999) treated material from the Murchison CM2 meteorite with $H_2O_2$ and found that it enhanced disaggregation of the particles because of its effect on the organic polymer fraction, which normally acts like cement. Progressive chemical oxidation of hydrocarbons by $H_2O_2$ accumulating within the interstices of a comet might also weaken adhesion within the nucleus accelerating its break-up.

Finally, it should be noted that if significant concentrations of $H_2O_2$ and/or $O_2$ accumulate and become mixed with hydrocarbons or CO, the possibility exists for a combustion reaction to be initiated. However, this author considers this to be a very unlikely event given that (a) an energetic ignition source would be required, and (b) repeat outbursts of the kind witnessed for 17P would not be expected.

## 3 INTERPRETATION OF THE 2007 OUTBURST OF COMET 17P/HOLMES

### 3.1 Disruption of the comet nucleus

As the comet approaches and passes perihelion, an increasing fraction of the subsurface $H_2O_2/H_2O$ ice will melt and can more easily percolate through any voids within the nucleus. Several months after perihelion in 2007, 17P may have accumulated a critical amount of aqueous $H_2O_2$ solution which then by chance reached a metal-rich region within the nucleus. The decomposition reaction to form oxygen gas then ensued, and being exothermic caused self-heating, which then melted more ice further accelerating the reaction. The in-situ production of gaseous $O_2$ along with associated volatilised hydrocarbons and CO may then have built up pressure rapidly, leading to a runaway expansion of the gas within a confined space thereby causing disruption of part of the comet's mantle and releasing large amounts of dust along with $O_2$, $H_2O$ and hydrocarbons into space.

Sekanina (CBET 1118) estimated the total mass of the dust cloud to be about $10^{11}$ kg assuming a bulk density of 1.5 g cm$^{-3}$. According to Snodgrass et al. (2006), the nucleus of 17P has an effective radius of 1.62 km. Thus the observed mass loss corresponds to about 0.5% of the total mass of the nucleus, and is similar to that of the talp or layer proposed by Belton et al. (2007) for the interior of Jupiter-family comet nuclei. By this mechanism, ejection of this mass into space would occur from the sunward face of the nucleus, and given its low strength (as shown by its subsequent disintegration), it would be expected to take place at relatively low speed, say at no more than 0.1 km s$^{-1}$ on average. At this speed, the ejected material would possess a total kinetic energy of about $5 \times 10^{14}$ J. To generate this amount of energy from $H_2O_2$ would require the decomposition of at least $1.4 \times 10^8$ kg$H_2O_2$ which, if present on average as a 5-%wt aqueous solution, would infer the liberation into space of at least an associated $4 \times 10^9$ kg of liquid $H_2O$. Any free liquid water would very quickly freeze in the vacuum of space since water vapour would have evaporated rapidly withdrawing latent heat (45.0 kJ mol$^{-1}$ at 273ºK) from the remainder and chilling it. Given that the enthalpy of fusion of $H_2O$ is 6.0 kJ mol$^{-1}$, some 13% or so of this liquid water would enter the gas phase. Such a cloud of water vapour would expand in an isotropic manner. For example, assuming an average temperature of 200ºK, a cloud exerting a pressure of 0.01 Torr would extend some 90 km in diameter.

**3.2 Development of the coma**

In addition to liquid water formed from catalytic decomposition of $H_2O_2$, a significant fraction of the material initially ejected into space would comprise water ice intimately mixed with dust grains and some volatile gases. If 20% of the solids comprised water ice, this represents an additional $25 \times 10^9$ kg of $H_2O$ ice. As the ejected material moved away from the nucleus, it would expand into the vacuum of space and the heat of the sun (solar irradiance = 230 W m$^{-2}$ at 2.44 AU) would begin to sublime the ice forming additional water vapour and leading to a further build-up of pressure within the cloud. During the first few hours, the cloud was so dense that it was opaque to sunlight falling upon it and as its cross-sectional area increased, it absorbed increasing amounts of heat from the sun resulting in sublimation of the ice to form free $H_2O$ molecules. Water within dust agglomerates would have also turned to vapour, blowing apart the friable mass of dust grains. If we assume a modest average expansion speed of 10 m s$^{-1}$, after 3 hours the cloud would have expanded to a sphere some 200 km in diameter. Given some 51.1 kJ/mol (at 240ºK) are required to sublime water and assuming that 10% of the incident radiation is used to sublime ice, the sublimation rate within a 200-km cloud would correspond to about 5% of the water ice per hour. The internal pressure within the cloud would further accelerate the dust creating a larger and larger cloud such that within a few more hours virtually all of the water ice would have sublimed away.

Whilst the water vapour continued to be produced, the temperature within the cloud would have remained low especially where the sun's radiation failed to penetrate. Once all of the ice had disappeared, the sun's energy would begin to raise the temperature of the dust grains and also, more importantly, the root-mean-square (rms) velocity of the free water molecules. So as the temperature of gaseous $H_2O$ began to increase, a further build-up of pressure would ensue leading to further acceleration of the dust, which would continue to expand in a spherically-symmetrical fashion. The maximum speed of expansion of the coma would be related to the effective temperature reached by the gaseous $H_2O$ within the cloud. For example, the rms velocity of $H_2O$ molecules at 180ºK is 500 m s$^{-1}$, i.e. similar to the expansion speed of the outer edge of the dust coma (CBET 1123).

In practice, the instantaneous velocity of individual $H_2O$ molecules comprises a Maxwell-Boltzmann distribution from zero upwards, with about 1% of the molecules moving at 1000 m s$^{-1}$ or slightly greater at a temperature of 180ºK. An outer faint bluish-green halo which developed after a few days (see http://aerith.net/comet/catalog/0017P/2007.html) is interpreted here as the tail-end of the velocity distribution of the $H_2O$ which escaped from within the dust cloud, expanded freely into space and was partially ionized by the solar wind.

### 3.3 Repeat comet outbursts

A reoccurrence of the outburst is easily possible via the mechanism proposed here. It depends on peroxide-rich water remaining within the subsurface where it is able to collect over a period of several weeks. It is important to recognise that the original 'explosion' is unlike that of a conventional fuel and oxidant which proceeds to a rapid conclusion once one of the reactants has been used up. Sudden decomposition of the aqueous peroxide is a localised event and much of the $H_2O_2$ remains intact after the initial outburst so that a repeat performance can take place several weeks later as observed in 1893. Further accumulation of aqueous $H_2O_2$, in-situ production of $O_2$ and self-heating may take many weeks before sufficient pressure is built up to cause a further disruption of the nucleus. The process is however an ad hoc one in that it depends on migration of the peroxide-rich water through voids to reach catalytically-active material buried within the nucleus.

### 3.4 Special nature of 17P/Holmes

The unique nature of the recent outburst of 17P does indicate that its composition and other properties must be atypical in some marked way. Based on the mechanism outlined here, it is postulated that the comet must be especially heterogeneous in composition with relatively large volumes being present where water remains isolated for long periods out of contact with catalytically-active minerals and other materials. In addition, a very slow rotation rate (<<1 rev d$^{-1}$) and very low albedo would accentuate the effect of continuous heating and cooling of the nucleus thereby helping the physical processes responsible for concentrating $H_2O_2$ within the subsurface. Indeed, the re-outburst of 17P in 1893 a little over 70 days after the first outburst might suggest a rotation period for the nucleus of about 75 ± 15 d opening up the possibility of an increase in activity of the comet in late December 2007 or soonafter. Alternatively, the rotation rate might equal 75/n ± 15/n d where n is a small integer, say 1, 2 or 3. Interestingly, the most notable example of a periodic comet exhibiting repeat outbursts is comet 29P/Schwassmann-Wachmann. Spitzer observations (Stansberry et al. 2004) have indicated from a study of the behaviour of the jets that this comet nucleus rotates with a period of 60 days or more, and that it has a very low geometric albedo of 0.025 ± 0.01. Both these characteristics strengthen the argument that the peroxide mechanism also underlies outbursts in 29P even though the comet remains some 6 AU distant from the Sun.

Regions rich in water may also contain dissolved materials which act to stabilise peroxide solutions at low concentration. Although the presence of trace inhibitors may stabilise $H_2O_2$, it is not a prerequisite of the proposed mechanism. In contrast to 17P, the majority of comets may have a much more homogeneous composition and so peroxide is catalytically decomposed close to the site of its formation preventing the accumulation of any significant quantity and limiting the scale of outbursts, if any, via this mechanism.

### 4 CONCLUSIONS

Although many mechanisms have been put forward to explain comet outbursts, these all appear to be somewhat lacking as an explanation of the recent outburst of 17P/Holmes in terms of the magnitude of the event, the fact that it has exhibited a repeat outburst over an interval of several months in the past, and that the nucleus of the comet remains intact. This study has for the first time highlighted the potential role of $H_2O_2$ in comet outbursts by drawing attention to certain important physico-chemical properties of the molecule, which taken as a whole provide a credible mechanism for explaining the outburst behaviour of 17P and possibly other short-period comets.

Only in recent years have observational methodologies permitted the direct detection of $H_2O_2$ mainly through IR studies from space-borne instruments. Ground-based observations are difficult because of the interference of $H_2O$ vapour and $O_2$ in our atmosphere. Also, the detection of $H_2O_2$ in the ejected material following an outburst would also be disguised both by the associated large excess of water vapour as well as the tendency for the molecule to decompose when exposed to ionising radiation in space. Given the new observational techniques now available, it is hoped that more attention will be paid not only to the study of water in comets but also to its oxidised analogue, $H_2O_2$.

Laboratory studies could also be used to test the conditions of $H_2O_2$ formation mooted here. In particular, the rate of formation on exposure of water/ice contained within a highly porous matrix to energetic particles typical of the solar wind and vacuum-UV radiation along with a simulation of the processes which may lead to its enrichment within the body of the nucleus would be valuable. Also, the catalytic activity for $H_2O_2$ decomposition could be usefully tested using a variety of meteorite-derived materials so as to verify the basis of this hypothesis.